\documentclass[12pt,letterpaper,copyedit]{article}
\usepackage{osajnl}
\usepackage{graphicx}
\usepackage{overcite,hyperref}

\begin{document}

\title{Interface localized modes and hybrid lattice solitons in waveguide arrays}

\author{Mario I. Molina}

\affiliation{Departamento de F\'{\i}sica, Facultad de Ciencias,
Universidad de Chile, Casilla 653, Santiago, Chile}

\author{Yuri S. Kivshar}

\affiliation{Nonlinear Physics Center, Research School of Physical
Sciences and Engineering, Australian National University, Canberra
ACT 0200, Australia}

\begin{abstract}
We discuss the formation of guided modes localized at the interface
separating two different periodic photonic lattices. Employing the
effective discrete model, we analyze linear and nonlinear interface
modes and also predict the existence of stable interface solitons
including the {\em hybrid staggered/unstaggered lattice solitons}
with the tails belonging to spectral gaps of different types.
\end{abstract}


\maketitle

\newpage

Surface modes have been studied in different branches of physics
including guided wave optics, where surface waves were predicted to
exist at interfaces separating periodic and homogeneous dielectric
media~\cite{Yeh_APL_78}. Recently, it was suggested
theoretically~\cite{OL_markis} and demonstrated
experimentally~\cite{PRL_suntsov} that nonlinearity-mediated
trapping of light near the edge of a truncated waveguide array with
self-focusing nonlinear response can lead to the formation of
nonlinear localized surface states which can be understood as
discrete optical solitons~\cite{book} localized near and trapped by
the surface~\cite{OL_our} for powers exceeding a certain
threshold value. One of the important generalizations of these ideas
is the concept of {\em multi-gap surface solitons}, i.e. mutually
trapped surface states with components associated with different
spectral gaps~\cite{OE_ivan}.

In this Letter, we study another important generalization of the
concept of nonlinear surface modes. We analyze linear and nonlinear
optical guided modes localized at an interface separating two
different semi-infinite periodic photonic lattices. In the framework
of an effective discrete model we demonstrate that the analysis of
linear interface states in such composite arrays provides an
important tool for analyzing the interface solitons and their basic
properties. We then find numerically the families of stable {\em
interface lattice solitons} including a novel class of {\em hybrid
staggered/unstaggered lattice solitons} with tails localized in
spectral gaps of different types.

We consider an interface separating two different semi-infinite
arrays of optical waveguides (as shown in the top of Fig.~1)
described by the system of coupled-mode equations~\cite{book} for
the normalized mode amplitudes $E_n$:
\begin{equation}
i \frac{dE_n}{d z} + \epsilon_n E_n + (E_{n+1} + E_{n-1}) + \gamma
|E_n|^2 E_n = 0,
\end{equation}
where the propagation coordinate $z$ is normalized to the intersite
coupling $V$, $E_n$ are defined in terms of the actual electric
field ${\cal E}_n$ as $E_n = (2V \lambda_0 \eta_0/\pi n_0
n_2)^{1/2}{\cal E}_n$, $\lambda_0$ is the free-space wavelength,
$\eta_0$ is the free-space impedance, $n_0$ and $n_2$ are the mean
values of the linear and nonlinear refractive indices of each
waveguide, and $\gamma$ defines the nonlinear response strength. The
waveguide interface is introduced by the condition: $\epsilon_0$ at
$n=0$, and $\epsilon_n = \epsilon_A$ or $\epsilon_{n}=\epsilon_B$
for negative or positive $n$, respectively.

First, we look for {\em linear surface modes} in the form $E_n = A\
\xi^{|n|}_{\pm} \exp (i \beta z)$ localized near the interface
waveguide with $n=0$, and obtain the condition $\xi_{+}/\xi_{-} =
\epsilon_{A0}/\epsilon_{B0}$ and the dispersion relation $ \beta =
\epsilon_{0} + \frac{1}{2}(\epsilon_{A0} + \epsilon_{B0})(1 -
\sqrt{1 + 4/\epsilon_{A0}\epsilon_{B0}})$, where
$\epsilon_{A0}\equiv \epsilon_{A}-\epsilon_{0}$ and
$\epsilon_{B0}\equiv \epsilon_{B}-\epsilon_{0}$. Figure~1 summarizes
our results for the existence of such localized states on the
parameter plane ($\epsilon_{A0}$, $\epsilon_{B0}$), as well as
displays examples of localized modes corresponding to different
existence regions. We note the existence of two sectors where no
localized states exist (shaded regions). One of them is bounded by
the curves $\epsilon_{B0} = 4/|\epsilon_{A0}|$ (for $-\infty <
\epsilon_{A0} < 0$) and $\epsilon_{B0} = \epsilon_{A0}/(1 -
\epsilon_{A0})$ (for $-\infty < \epsilon_{A0} < 1$). The other one
is obtained by a reflection across the origin. Inside these regions
either $|\xi_{+1}|$, $|\xi_{-1}|$ or both exceed one.

One of the important observations that follows from our analysis is
the existence of {\em hybrid staggered/unstaggered interface modes}
for the opposite sign of the propagation constant mismatches of two
lattices. These modes have the tails localized in the bandgaps of
different types, i.e. above (for one array) and below (for the other
array) of the spectral band.

%%%%%%%%%%%%%%%%%%%%%%%%
\begin{figure}[t]
\begin{center}
\includegraphics[width=2.8in]{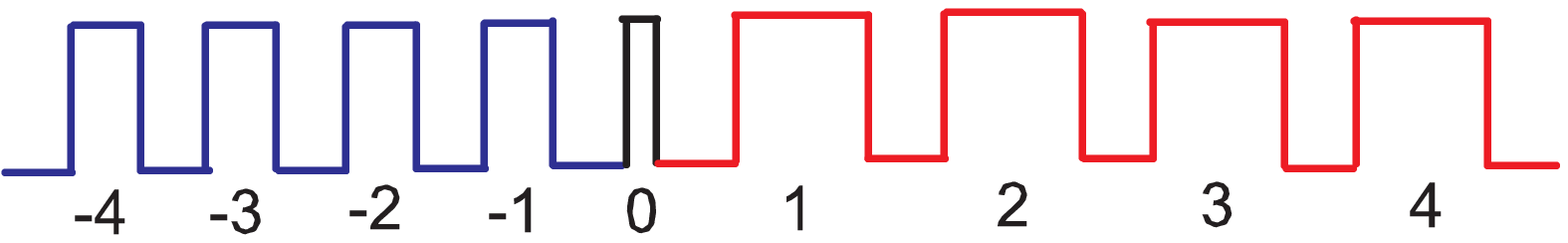}\\
\includegraphics[width=3.0in]{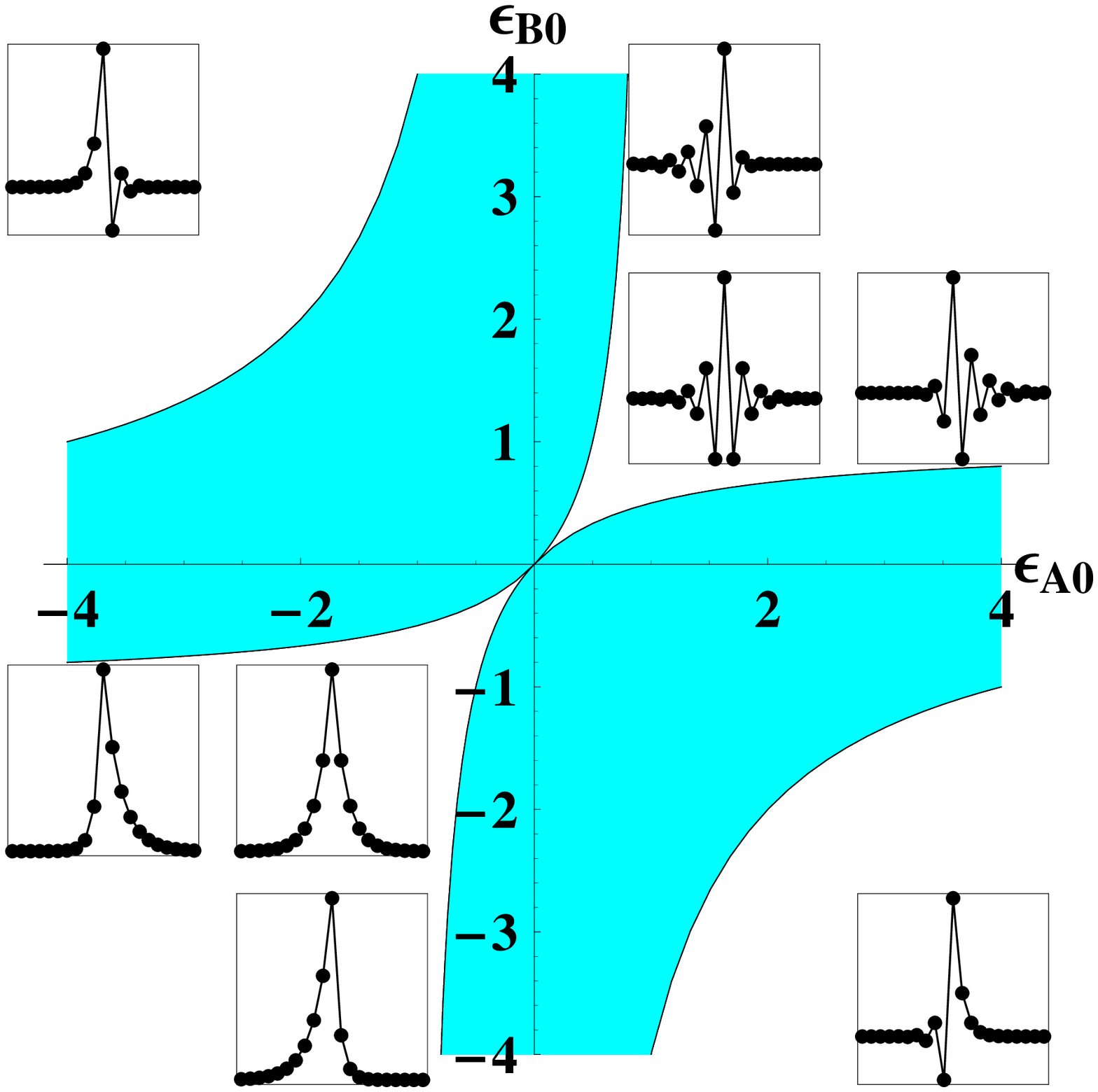}
\caption{(Color online) Phase diagram of different types of
localized interface modes. No localized modes exist inside the
shaded areas. The insets show examples of localized modes
corresponding to different values of $\epsilon_{A0}\equiv
\epsilon_{A}-\epsilon_{0}$ and $\epsilon_{B0}\equiv
\epsilon_{B}-\epsilon_{0}$. Top: schematic structure of the
waveguide interface. } \label{fig1}
\end{center}
\end{figure}
%%%%%%%%%%%%%%%%%%%%%%%

In Figs.~2(a,b) we show the propagation constant $\beta$ of the
localized modes as a function of the interface parameter
$\epsilon_{0}$, for the characteristic cases (a) $\epsilon_{A} =
0.6, \epsilon_{B} = -0.6$, and (b)  $\epsilon_{A} = 3, \epsilon_{B}
= -3$. We note that the mode always lies outside the linear spectral
bands, but its structure depends strongly on the overlap of the
bands, so that the hybrid modes appear for a relatively large band
mismatch, as shown in Fig.~2(b) (middle curve).

%%%%%%%%%%%%%%%%%%%%%%%
\begin{figure}[t]
\begin{center}
\includegraphics[width=3.5in]{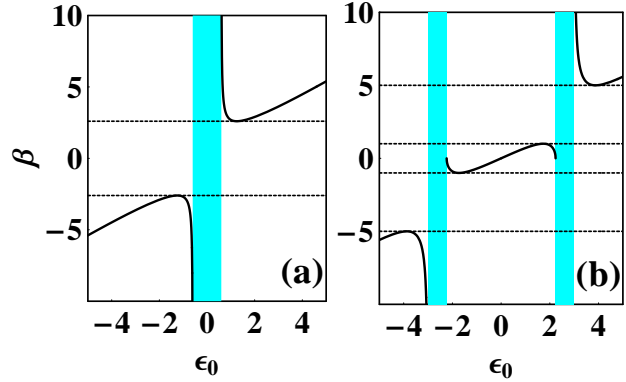}
\caption{(Color online) Families of localized modes shown as the
dependencies $\beta$ vs. $\epsilon_{0}$, for the cases: (a)
$\epsilon_{A} = 0.6, \epsilon_{B} = -0.6$, and (b) $\epsilon_{A} =
3, \epsilon_{B} = -3$. No localized modes exist inside the shaded
regions. The dashed lines mark the spectral bands of both arrays,
which in the case (b) do not overlap with each other. }
\label{fig2}
\end{center}
\end{figure}
%%%%%%%%%%%%%%%%%%%%%%%%%%

The analysis of linear localized interface modes in such an array
provides an important information about the existence of nonlinear
interface modes---{\em lattice surface solitons}. Next, we consider
two semi-infinite nonlinear waveguide arrays characterized by 
propagation constants $\epsilon_{A}$ and $\epsilon_{B}$ that are
joined by an interface waveguide with the propagation constant
$\epsilon_{0}$. We focus on the interface modes defined by having
their centers at either the first of the A waveguides or the first
of the B waveguides. We find different classes of nonlinear
localized modes and study their linear stability numerically via a
multidimensional Newton-Raphson method.

%%%%%%%%%%%%%%%%%%%%%%%
\begin{figure}[t]
\begin{center}
\includegraphics[width=4.0in]{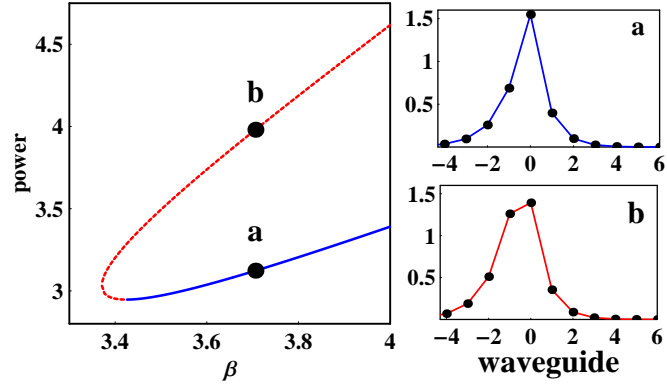}
\caption{(Color online) Power vs. propagation constant for the
interface unstaggered solitons centered on the first A waveguide
(for $\epsilon_{A}=\epsilon_{0}=0.6$, $\epsilon_{B}=-0.6$). The
solid (dashed) curve refers to the stable (unstable) branches.
Inserts show two examples of the interface modes.} \label{fig3}
\end{center}
\end{figure}
%%%%%%%%%%%%%%%%%%%%%%%%%%

First, we consider the case $\epsilon_{0} = \epsilon_{A} =
-\epsilon_{B}= 0.6$ and $\gamma = +1$. In the linear limit, this
corresponds to a negative vertical line in Fig.~1, where no
localized modes exist. The presence of nonlinearity shifts the
propagation constant of the mode to the left, until it gives rise to
an unstaggered localized mode. Therefore, we predict the lowest
nonlinear interface mode to be unstaggered. This is indeed confirmed
by our numerical computations, and the family of the lowest-order
interface nonlinear modes is shown in Fig.~3, where the upper/lower
branch corresponds to unstable/stable modes. Next, we consider the
case $\epsilon_{A}=3$, $\epsilon_{0}=0$ and $\epsilon_{B}=-3$.
Results are summarized in Fig. 4 which shows the dependence of the
power vs. propagation constant for several low-power modes. We note
that the lowest mode extends all the way to zero power, and
therefore it corresponds in that limit to the linear  mode induced
by three concurrent dissimilar propagation constants. More
importantly, the mode amplitudes show now a {\em hybrid character},
being unstaggered in one side of the interface and staggered on the
other.

In addition, we find many other types of nonlinear interface modes
including twisted and flat-top modes, as well as the modes with
different location of their centers relative to the interface, as
discussed earlier for a semi-infinite waveguide array~\cite{OL_our}.
As a special limit of those modes, we find also the {\em kink
surface modes} which are extended in one direction while being localized
in the other.

%%%%%%%%%%%%%%%%%%%%%%%
\begin{figure}[t]
\begin{center}
\includegraphics[width=3.0in]{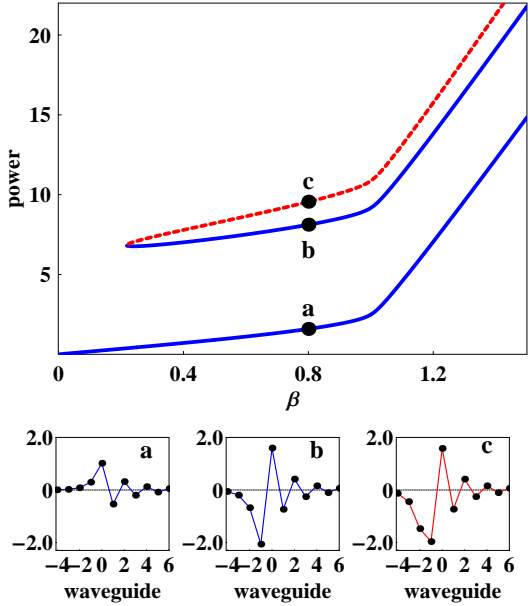}
\vspace{-0.75in}
\caption{(Color online) Power vs. propagation
constant for the hybrid interface staggered/unstaggered lattice
solitons for $\epsilon_{A}=-3, \epsilon_{0}=0, \epsilon_{B}=3$.
Solid (dashed) curves refer to the stable (unstable) branches.
Inserts show three examples of the hybrid interface solitons. }
\label{fig4}
\end{center}
\end{figure}
%%%%%%%%%%%%%%%%%%%%%%%%%%

Next, we study the evolution of all types of nonlinear interface
modes checking in this way our stability results and analyzing the
evolution scenario for the unstable modes. In all cases examined, we
observe that the unstable modes decay into the unstaggered
fundamental 

%%%%%%%%%%%%%%%%%%%%%%%
\begin{figure}[thb]
\begin{center}
\includegraphics[width=2.75in]{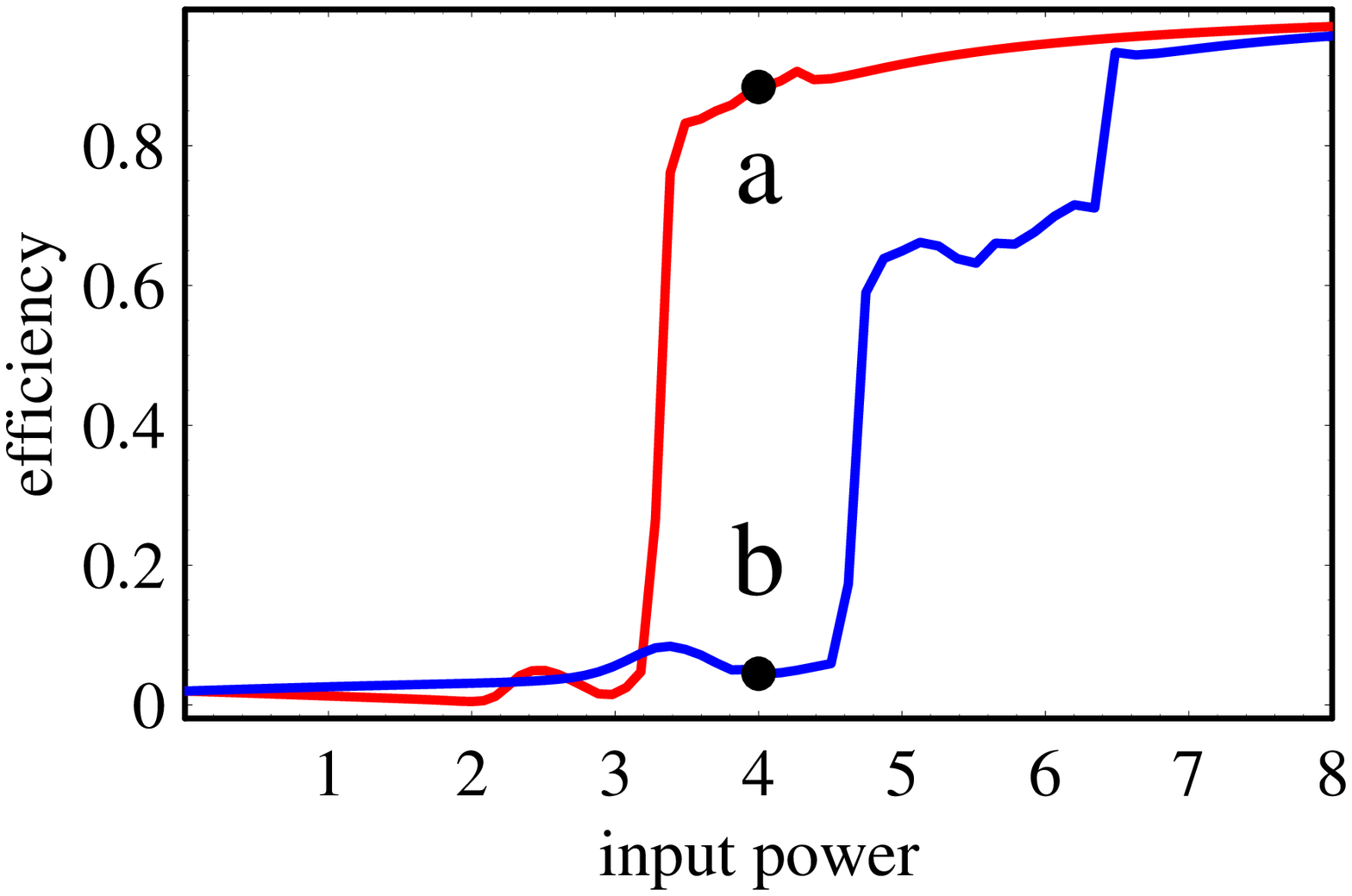}\\
\includegraphics[width=2.75in]{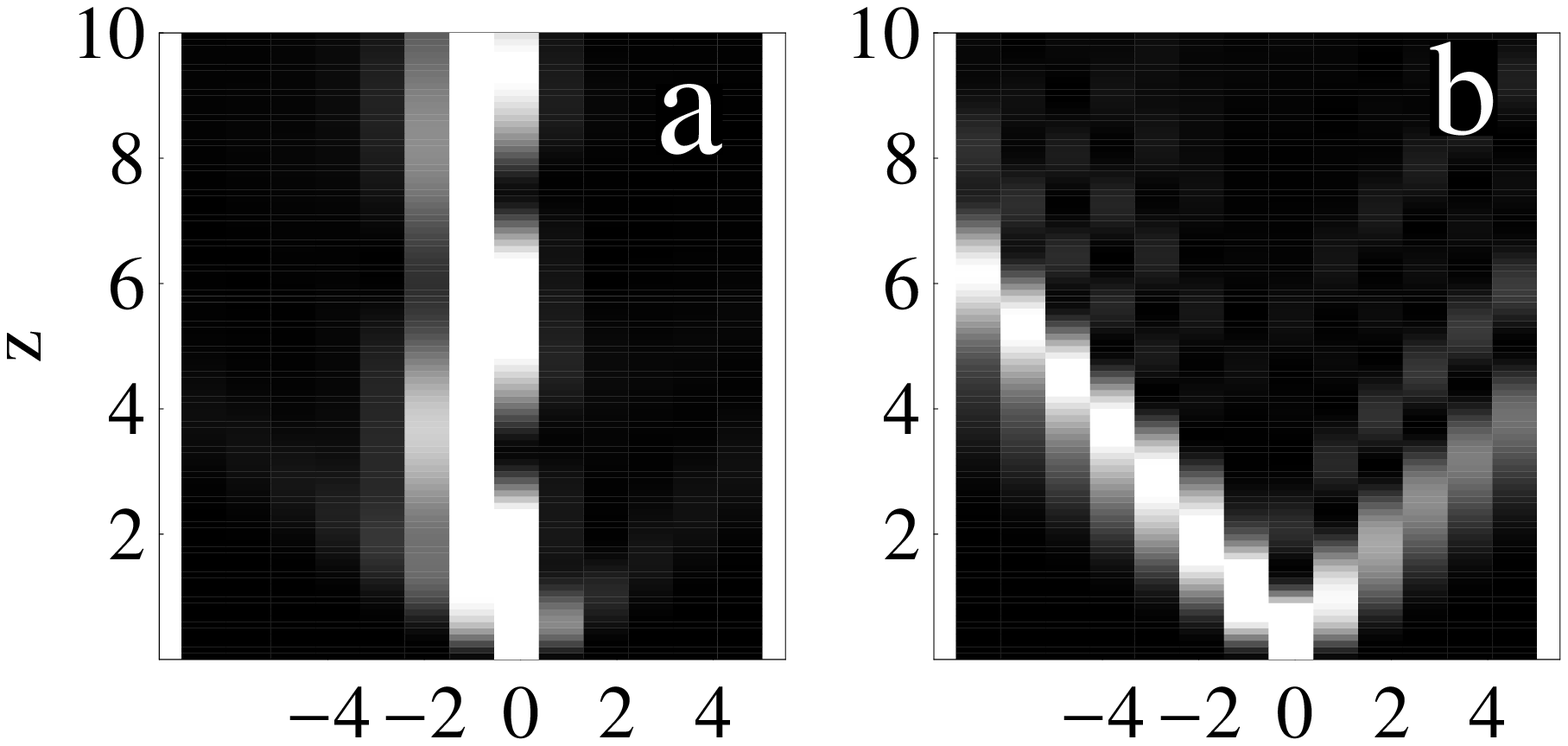}
\caption{(Color online) Top: trapping efficiency for the
generation of interface solitons. The red (blue) curve denotes the
case when the initial input is centered on the first A (B)
waveguide. Bottom: evolution of initial states marked `a' and `b'
(for $\epsilon_{A}=0.6$, $\epsilon_{B} =-0.6$).} \label{fig5}
\end{center}
\end{figure}
%%%%%%%%%%%%%%%%%%%%%%%%%%

mode by emitting radiation. In particular, for the
unstaggered unstable modes, this decay is rapid and it proceeds
vertically. For other cases, the unstable localized modes decay into
higher-power generalizations of the fundamental mode. Also, we
observe that unstable unstaggered modes decay much faster than
unstable twisted, flat-top, or dark-like nonlinear modes.

Finally, we study how an interface lattice soliton can be generated
in experiment by exciting a single waveguide which is either the
very first of the A waveguides, or the very first of the B
waveguides. We define the trapping efficiency as the power fraction
$P_{\rm out}/P_{\rm in}$ remaining in the $10$ central waveguides.
Results for the trapping efficiency are shown in Fig.~5(top), where
we have used $201$ waveguides, a total evolution length of $20$. The
bottom portion of Fig. 5 shows the evolution of the initial states
marked by the points `a' and `b' in Fig.~5(top).

We would like to emphasize that the results obtained here can be
easily generalized to the case of {\em surface gap solitons} which
were predicted theoretically~\cite{prl_torner} and observed
experimentally~\cite{prl_exp} in periodic photonic lattices with
defocusing nonlinearity, where surface solitons appear in the gaps
of the photonic bandgap spectra or their overlaps.

In conclusion, we have analyzed different types of linear and
nonlinear optical guided modes localized at the interface separating
two different semi-infinite periodic photonic lattices. In the
framework of an effective discrete nonlinear model, we have
demonstrated the existence of stable interface lattice solitons
including the hybrid staggered/unstaggered discrete solitons with
the tails that belong to different spectral gaps. We believe our
results will encourage the first experimental observations of this
novel type of surface optical solitons in photonic lattices.

This work has been supported by Conicyt and Fondecyt grants 1050193
and 7050173 in Chile, and by the Australian Research Council in
Australia.

\newpage

\section*{\centerline{List of Figure Captions}}

\noindent
Figure 1:\ \ (Color online) Phase diagram of different types of
localized interface modes. No localized modes exist inside the
shaded areas. The insets show examples of localized modes
corresponding to different values of $\epsilon_{A0}\equiv
\epsilon_{A}-\epsilon_{0}$ and $\epsilon_{B0}\equiv
\epsilon_{B}-\epsilon_{0}$. Top: schematic structure of the
waveguide interface.
\vspace{0.4cm}

\noindent
Figure 2:\ \ (Color online) Families of localized modes shown as the
dependencies $\beta$ vs. $\epsilon_{0}$, for the cases: (a)
$\epsilon_{A} = 0.6, \epsilon_{B} = -0.6$, and (b) $\epsilon_{A} =
3, \epsilon_{B} = -3$. No localized modes exist inside the shaded
regions. The dashed lines mark the spectral bands of both arrays,
which in the case (b) do not overlap with each other.
\vspace{0.4cm}

\noindent
Figure 3:\ \ (Color online) Power vs. propagation constant for the
interface unstaggered solitons centered on the first A waveguide
(for $\epsilon_{A}=\epsilon_{0}=0.6$, $\epsilon_{B}=-0.6$). The
solid (dashed) curve refers to the stable (unstable) branches.
Inserts show two examples of the interface modes.
\vspace{0.4cm}

\noindent
Figure 4:\ \ (Color online) Power vs. propagation
constant for the hybrid interface staggered/unstaggered lattice
solitons for $\epsilon_{A}=-3, \epsilon_{0}=0, \epsilon_{B}=3$.
Solid (dashed) curves refer to the stable (unstable) branches.
Inserts show three examples of the hybrid interface solitons.
\vspace{0.4cm}

\noindent
Figure 5:\ \ (Color online) Top: trapping efficiency for the
generation of interface solitons. The red (blue) curve denotes the
case when the initial input is centered on the first A (B)
waveguide. Bottom: evolution of initial states marked `a' and `b'
(for $\epsilon_{A}=0.6$, $\epsilon_{B} =-0.6$).

\end{document}